\documentclass[12pt,a4paper]{article}
\usepackage{amsfonts}
\usepackage{amssymb}
\usepackage{amsmath}
\usepackage{latexsym}
\newtheorem{trm}{Theorem}
\newtheorem{col}{Corollary}
\newtheorem{prop}{Proposition}
\begin{document}

\begin{center}
{\Large \bf

Non-existence of non-topological solitons in some types\\
\vspace{2mm} of gauge field theories in Minkowski space} \\

\vspace{4mm}

Mikhail N. Smolyakov\\
\vspace{0.5cm} Skobeltsyn Institute of Nuclear Physics, Moscow
State University,
\\ 119991, Moscow, Russia\\
\end{center}

\begin{abstract}
In this paper the conditions, under which non-topological solitons
are absent in the Yang-Mills theory coupled to a non-linear scalar
field in Minkowski space, are obtained in a very simple way. It is
also shown that non-topological solitons are absent in the theory
describing the massive complex vector field coupled to the
electromagnetic field in Minkowski space.
\end{abstract}

\section{Introduction}
In this paper we present some non-existence results for gauge
field theories, which can be obtained with the help of the
so-called "scaling arguments". These scaling arguments are based
on the use of an appropriate form of the variations of fields. For
the first time this technique was used in \cite{Hobart,Derrick} to
show the absence (or to find necessary conditions for existence,
see \cite{Rosen1}) of solitons in a non-linear scalar field theory
(see also discussions in \cite{Rajaraman,Rubakov}). Later such
methods were used in models with more complicated static
configurations of fields, for example, in the case of Yang-Mills
field coupled to a scalar field \cite{Rubakov} and for skyrmions,
monopoles and instantons \cite{Dimopolous,Krasnikov,Manton}, as
well as in models admitting time-independent effective Lagrangians
\cite{Makhankov}.

In this paper we will use the scaling arguments to show the
absence of non-topological solitons (when all the fields vanish at
spatial infinity) in some types of gauge field theories in
Minkowski space even if configurations of fields are
time-dependent. We restrict ourselves to the case when solutions
to equations of motion are periodic in time with a period
$T<\infty$, but not necessarily of the simplest form $\sim
e^{i\omega t}$ (it should be noted, that the scaling arguments
were used long ago in \cite{Rosen1} for obtaining restrictions on
the existence of soliton solutions periodic in time in a
non-linear scalar field theory). This restriction is necessary to
make the proof mathematically rigorous, below we will discuss it
in more detail. We will obtain the conditions for the scalar field
potential which ensure the absence of non-topological solitons in
a theory describing Yang-Mills field coupled to a non-linear
scalar field (analogous scaling arguments were applied to the case
of Klein-Gordon-Maxwell system with some simplifications, see
\cite{Smolyakov}). For non-negative scalar field potentials we get
a restriction, which is in agreement with that obtained in
\cite{GS} using a different technique. Another consequence of our
analysis is the absence of non-topological solitons in pure
Yang-Mills theory, which is of course a well-known result and was
obtained in \cite{Deser}-\cite{GS1} also using different
techniques. Finally we will apply our methods to the theory of
U(1)-charged massive vector field to show the absence of
non-topological solitons in this case.

\section{Yang-Mills field coupled to a non-linear scalar field}
Let us consider the following form of the four-dimensional action:

\begin{equation}\label{act}
S=\int
d^{4}x\left[\eta^{\mu\nu}(D_{\mu}\phi)^{\dagger}D_{\nu}\phi-V(\phi^{\dagger}\phi)-\frac{1}{4}{F^{a}}^{\mu\nu}F_{\mu\nu}^{a}\right],
\end{equation}
where $\eta_{\mu\nu}=diag(1,-1,-1,-1)$ is the flat Minkowski
metric,
\begin{equation}
F_{\mu\nu}^a=\partial_{\mu}A_{\nu}^{a}-\partial_{\nu}A_{\mu}^{a}+gC^{abc}A_{\mu}^{b}A_{\nu}^{c},
\end{equation}
\begin{equation}
D_{\mu}\phi=\partial_{\mu}\phi-igT^{a}A_{\mu}^{a}\phi,
\end{equation}
where $C_{abc}$ are the structure constants of a compact gauge
group and $T^{a}$ are generators of the group in the
representation space of field $\phi$. We also suppose that
\begin{equation}\label{vpotgeneral}
V(\phi^{\dagger}\phi)|_{\phi^{\dagger}\phi=0}=0,\quad
\frac{dV(\phi^{\dagger}\phi)}{d(\phi^{\dagger}\phi)}\Biggr|_{\phi^{\dagger}\phi=0}=C,\quad
|C|<\infty.
\end{equation}
The latter condition ensures that the trivial solution is
$\phi\equiv 0$, $A_{\mu}=gT^{a}A_{\mu}^{a}\equiv 0$. The condition
$V(\phi^{\dagger}\phi)|_{\phi^{\dagger}\phi=0}=0$ is imposed for
simplicity, in order not to deal with additive constants. We also
suppose that:
\begin{enumerate}
\item
there are no sources which are external to the system described by
action (\ref{act});
\item
solutions to equations of motion are periodic in time with a
period $\tilde T<\infty$ up to a coordinate shift and a spatial
rotation, i.e. for all fields on the solution the relation
$\Psi(t+\tilde T,\vec
x)\equiv\Lambda(\Omega)\Psi(t,\Omega^{-1}\vec x-\vec l)$ must hold
for any $t$, where $\Psi(t,\vec x)$ schematically represents the
field under consideration, $\vec l$ is a constant vector,
$\Omega\in SO(3)$ and $\Lambda(\Omega)$ denotes the representation
of the rotation group carried by the field $\Psi(t,\vec x)$ (this
means that we also consider solitons which may not be at rest and
spinning solitons).
\end{enumerate}
Let us discuss the second assumption. It is not difficult to show
that the periodicity condition presented above corresponds to a
rotation in the plane orthogonal to some axis of rotation and a
rectilinear motion along this axis. For this reason, one can
always choose a suitable coordinate system in which the axis of
rotation passes through the point $\vec x=0$ and $\vec l\,$
belongs to this axis, i.e. $\Omega\vec l=\vec l$. In this
coordinate system the velocity of the rectilinear motion is $\vec
l/{\tilde T}$ and we suppose that $|\vec l|/{\tilde T}<1$ (i.e.
smaller than the speed of light). Next, with the help of a Lorentz
transformation we can always pass to the coordinate system where
$\vec l=0$. In this new coordinate system $\Psi(t+T,\vec
x)\equiv\Lambda(\Omega)\Psi(t,\Omega^{-1}\vec x)$ with a new
period $T$ and we can rewrite the initial action as
\begin{eqnarray}\label{rotateff}
S=\int_{-\infty}^{\infty} dt\int d^{3}x L[\Psi(t,\vec x)]=
\sum_{n=-\infty}^{\infty}\int_{nT}^{(n+1)T} dt\int d^{3}x
L[\Psi(t,\vec x)]=\\ \nonumber
=\sum_{n=-\infty}^{\infty}\int_{0}^{T} dt\int d^{3}x
L[\Psi(t+nT,\vec x)]=\sum_{n=-\infty}^{\infty}\int_{0}^{T} dt\int
d^{3}x L[\Lambda^{n}(\Omega)\Psi(t,\Omega^{-n}\vec x)]=\\
\nonumber=\sum_{n=-\infty}^{\infty}\int_{0}^{T} dt\int d^{3}x
L[\Psi(t,\vec x)],
\end{eqnarray}
where we have made change of variables $\vec x\to \Omega^{n}\vec
x$ in each integral. Thus, we can use the effective action
\begin{eqnarray}\label{effact001}
S_{eff}=\int_{0}^{T} dt\int d^{3}x L[\Psi(t,\vec x)]
\end{eqnarray}
instead of the initial one. We will discuss the necessity of our
restriction to consider such types of solutions at the end of this
section.

It is interesting to note that if we suppose that for a solution
there is no $N\ne 0$ such that $\Omega^{N}=1$, where $N$ is
integer, then this solution will not be periodic in the usual
sense, but nevertheless we can use the effective action of form
(\ref{effact001}) for examining such types of solitons.

Now let us proceed to the system described by action (\ref{act}).
The corresponding effective action takes the form (here and below
we omit the subscript "\textit{eff}" for effective actions)
\begin{eqnarray}\label{effact1}
S=\int_{0}^{T} dt\int
d^{3}x\left[(D_{0}\phi)^{\dagger}D_{0}\phi-(D_{i}\phi)^{\dagger}D_{i}\phi-V(\phi^{\dagger}\phi)+\right.\\
\nonumber \left.+\frac{1}{2}F_{0i}^{a}F_{0i}^{a}-
\frac{1}{4}F_{ij}^{a}F_{ij}^{a}\right],
\end{eqnarray}
where $i,j=1,2,3$. Let us denote
\begin{eqnarray}\label{p1}
\int_{0}^{T}dt\int d^{3}x
(D_{0}\phi)^{\dagger}D_{0}\phi=\Pi_{0}\ge 0,\\ \label{p2}
\int_{0}^{T}dt\int d^{3}x
(D_{i}\phi)^{\dagger}D_{i}\phi=\Pi_{1}\ge 0,\\ \label{p3}
\int_{0}^{T}dt\int d^{3}x \frac{1}{2}F_{0i}^{a}F_{0i}^{a}=\Pi_{A0}\ge 0,\\
\label{pn} \int_{0}^{T}dt\int d^{3}x
\frac{1}{4}F_{ij}^{a}F_{ij}^{a}=\Pi_{A1}\ge 0.
\end{eqnarray}
We suppose that all these integrals are finite.

We will be looking for smooth solutions to equations of motion,
following from (\ref{act}), with the asymptotic form
\begin{equation}\label{sol1}
\lim\limits_{x^{i}\to\pm\infty}\phi(t,\vec x)=0,
\end{equation}
\begin{equation}\label{sol2}
\lim\limits_{x^{i}\to\pm\infty}A_{\mu}(t,\vec x)=0.
\end{equation}

\begin{trm}
For potentials satisfying (\ref{vpotgeneral}) non-topological
solitons of form (\ref{sol1}), (\ref{sol2}), periodic in time up
to a spatial rotation and a coordinate shift such that $\Omega\vec
l=\vec l$, $|\vec l|/{\tilde T}<1$, with integrals
(\ref{p1})-(\ref{pn}) and integrals $\int_{0}^{T}dt\int
d^3x\frac{dV(\phi^{\dagger}\phi)}{d(\phi^{\dagger}\phi)}\phi^{\dagger}\phi$,
$\int_{0}^{T}dt\int d^3xV(\phi^{\dagger}\phi)$ finite, are absent
in the theory with action (\ref{act}) if there exists $\gamma:\,
\frac{1}{2}<\gamma\le\frac{3}{2}$ for which the inequality
\begin{equation}\label{theor1}
2\gamma\frac{dV(\phi^{\dagger}\phi)}{d(\phi^{\dagger}\phi)}\phi^{\dagger}\phi-3V(\phi^{\dagger}\phi)\ge
0
\end{equation}
is fulfilled for any $\phi$ (or at least for the range of values
of the field $\phi$, which is supposed to occur in the solution).
\end{trm}
{\bf Proof:}\\
Let us suppose that there exists a solution $\phi(t,\vec x)$,
$A_{\mu}(t,\vec x)$. With the help of (\ref{p1})-(\ref{pn}) we get
from (\ref{effact1})
\begin{eqnarray}
S=\Pi_{0}-\Pi_{1}-\int_{0}^{T}dt\int d^3x
V\left(\phi^{\dagger}\phi\right)+\Pi_{A0}- \Pi_{A1}.
\end{eqnarray}
Now let us consider the following transformation of our solution
\begin{eqnarray}\label{resc1}
\phi(t,\vec x)\to\lambda^{\gamma}\phi(t,\lambda\vec x),\\
\label{resc2} A_{0}^{a}(t,\vec x)\to A_{0}^{a}(t,\lambda\vec x),\\
\label{resc3} A_{i}^{a}(t,\vec x)\to \lambda
A_{i}^{a}(t,\lambda\vec x)
\end{eqnarray}
with a real parameter $\lambda$. The action on this transformed
solution takes the form
\begin{eqnarray}
S=\lambda^{2\gamma-3}\Pi_{0}-\lambda^{2\gamma-1}\Pi_{1}-\\
\nonumber -\lambda^{-3}\int_{0}^{T}dt\int d^3x
V\left(\lambda^{2\gamma}\phi^{\dagger}(t,\vec x)\phi(t,\vec
x)\right)+\lambda^{-1}\Pi_{A0}- \lambda\Pi_{A1}.
\end{eqnarray}
Since we suppose that $\phi(t,\vec x)$, $A_{\mu}(t,\vec x)$ is a
solution to the equations of motion, the variation of the action
on this solution should vanish for any variations of the fields.
For the case of the transformations described by
(\ref{resc1})-(\ref{resc3}) it means that
\begin{eqnarray}\label{variation}
\frac{dS}{d\lambda}|_{\lambda=1}=(2\gamma-3)\Pi_{0}-(2\gamma-1)\Pi_{1}-\\
\nonumber -\int_{0}^{T}dt\int
d^3x\left(2\gamma\frac{dV(\phi^{\dagger}\phi)}{d(\phi^{\dagger}\phi)}\phi^{\dagger}\phi-3V(\phi^{\dagger}\phi)\right)
-\Pi_{A0}-\Pi_{A1}=0.
\end{eqnarray}
Indeed, $\lambda=1$ in (\ref{resc1})-(\ref{resc3}) corresponds to
the solution.

Now we are ready to consider the consequences following from
equation (\ref{variation}).
\begin{enumerate}
\item
$\frac{1}{2}<\gamma<\frac{3}{2}$. If
\begin{equation}
2\gamma\frac{dV(\phi^{\dagger}\phi)}{d(\phi^{\dagger}\phi)}\phi^{\dagger}\phi-3V(\phi^{\dagger}\phi)\ge
0
\end{equation}
for any $\phi$, then $\Pi_{0}=\Pi_{1}=\Pi_{A0}=\Pi_{A1}\equiv 0$
($2\gamma\frac{dV(\phi^{\dagger}\phi)}{d(\phi^{\dagger}\phi)}-3V(\phi^{\dagger}\phi)=0$
also), in this case $F_{\mu\nu}^a\equiv 0$ (this equality means
that $A_{\mu}$ is a pure gauge and we can set $A_{\mu}\equiv 0$).
From $\Pi_{0}=\Pi_{1}\equiv 0$ with $A_{\mu}\equiv 0$ and
according to (\ref{sol1}) we get $\phi\equiv 0$. Thus, solitons of
form (\ref{sol1}), (\ref{sol2}) are absent in the theory.
\item
$\gamma=\frac{3}{2}$. If
\begin{equation}\label{ineqq1}
\frac{dV(\phi^{\dagger}\phi)}{d(\phi^{\dagger}\phi)}\phi^{\dagger}\phi-V(\phi^{\dagger}\phi)\ge
0
\end{equation}
for any $\phi$, then $\Pi_{1}=\Pi_{A0}=\Pi_{A1}\equiv 0$, in this
case $A_{\mu}\equiv 0$, $\phi=\phi(t)\equiv 0$ (see (\ref{sol1}))
and solitons of form (\ref{sol1}), (\ref{sol2}) are also absent in
the theory.
\item
$\gamma=\frac{1}{2}$. If
\begin{equation}
\frac{dV(\phi^{\dagger}\phi)}{d(\phi^{\dagger}\phi)}\phi^{\dagger}\phi-3V(\phi^{\dagger}\phi)\ge
0
\end{equation}
for any $\phi$, then $\Pi_{0}=\Pi_{A0}=\Pi_{A1}\equiv 0$, in this
case $A_{\mu}\equiv 0$, $\phi=\phi(\vec x)$. To obtain
restrictions for this case, we can use the results of the
well-known Derrick theorem \cite{Derrick}, which states that in
the scalar field theories described by the standard action static
solitons are absent if $V(\phi^{\dagger}\phi)\ge 0$. Thus,
solitons of form (\ref{sol1}), (\ref{sol2}) are absent in the
theory if
\begin{equation}\label{eq24}
\frac{dV(\phi^{\dagger}\phi)}{d(\phi^{\dagger}\phi)}\phi^{\dagger}\phi\ge
3V(\phi^{\dagger}\phi)\ge 0
\end{equation}
for any $\phi\ne 0$. But the latter inequality is more stringent
than that following from (\ref{ineqq1}) for
$V(\phi^{\dagger}\phi)\ge 0$ (i.e. (\ref{eq24}) covers a narrower
range of potentials).
\end{enumerate}
{\bf End of the proof.}\\
{\bf Remark:} When considering static configuration of fields we
can also take the following transformations of the fields:
\begin{eqnarray}\label{rescmm}
\phi(t,\vec x)=\phi(\vec x)\to\lambda^{\gamma}\phi(\vec x),\\
\label{rescmm1} A_{0}^{a}(t,\vec x)=A_{0}^{a}(\vec x)\to
\lambda^{\beta}A_{0}^{a}(\vec x),\\
\label{rescmm2} A_{i}^{a}(t,\vec x)=A_{i}^{a}(\vec x)\to
A_{i}^{a}(\vec x)
\end{eqnarray}
with $\gamma>0$, $\beta<-\gamma$. Then we get
\begin{eqnarray}
\frac{dS^{\phi}}{d\lambda}|_{\lambda=1}=2(\gamma+\beta)\Pi_{0}-\\
\nonumber -2\gamma\left[\Pi_{1}+\int_{0}^{T}dt\int
d^3x\frac{dV(\phi^{\dagger}\phi)}{d(\phi^{\dagger}\phi)}\phi^{\dagger}\phi\right]+2\beta\Pi_{A0}=0.
\end{eqnarray}
Thus if
\begin{eqnarray}\label{statabs}
\frac{dV(\phi^{\dagger}\phi)}{d(\phi^{\dagger}\phi)}\phi^{\dagger}\phi\ge
0,
\end{eqnarray}
then $\Pi_{0}=\Pi_{1}=\Pi_{A0}\equiv 0$. Relation $\Pi_{A0}\equiv
0$ implies $A_{0}\equiv 0$, from $\Pi_{1}\equiv 0$ it follows that
$D_{i}\phi\equiv 0$ and thus
$\partial_{i}(\phi^{\dagger}\phi)\equiv 0$, which implies
$\phi\equiv 0$ (see (\ref{sol1})). Then it is very easy to show
that $A_{i}\equiv 0$, for example, applying (\ref{resc3}) to the
action containing $A_{i}$ only. Thus, we get the absence of static
solitons if condition (\ref{statabs}) is fulfilled. This
restriction was previously obtained in \cite{Malec} (see also
\cite{Malec1}).
\begin{col}
\begin{enumerate}
\item
For $V(\phi^{\dagger}\phi)\ge 0$, solitons of form (\ref{sol1}),
(\ref{sol2}) are absent if
\begin{equation}\label{color1}
\frac{dV(\phi^{\dagger}\phi)}{d(\phi^{\dagger}\phi)}\phi^{\dagger}\phi-V(\phi^{\dagger}\phi)\ge
0.
\end{equation}
\item
For $V(\phi^{\dagger}\phi)\le 0$, solitons (even unstable) of form
(\ref{sol1}), (\ref{sol2}) are absent if
\begin{equation}
\frac{dV(\phi^{\dagger}\phi)}{d(\phi^{\dagger}\phi)}\phi^{\dagger}\phi-3V(\phi^{\dagger}\phi)>
0.
\end{equation}
\item The restrictions presented above are valid for the models
with the scalar field only, i.e. if we drop the gauge field from
the theory.
\item Non-topological solitons satisfying the conditions presented
above are absent in the pure Yang-Mills theory.
\end{enumerate}
\end{col}
The proof of this corollary follows directly from Theorem~1.\\
{\bf Remark:} for the first time the no-go condition
(\ref{color1}) for $V(\phi^{\dagger}\phi)\ge 0$ and the potentials
$V(\phi^{\dagger}\phi)$ including a positive mass term was
obtained in a different way in \cite{GS} (for the case of a single
complex scalar field coupled to the electromagnetic field this
condition was obtained in \cite{Rosen2}). The absence of solitons
("classical lumps") in the pure Yang-Mills theory was shown long
time ago for the static \cite{Deser}, periodic \cite{Pagels} and
general cases \cite{Coleman1,Coleman2,Weder,Magg,GS1} (the
corresponding conjecture was discussed earlier, see, for example
\cite{Rosen3} and references therein).

It should be noted that the results presented above can be
obtained by taking the equations of motion following from action
(\ref{act}), multiplying them by variations of fields following
from (\ref{resc1})-(\ref{resc3}) (or from
(\ref{rescmm})-(\ref{rescmm2})), integrating over the four-volume
and combining the results coming from different equations of
motion (as an example see \cite{Rosen1}, where such an alternative
way was presented). Indeed, the variational principle gives us the
equations of motion and thus all the results obtained above can be
also obtained with the help of the equations of motion. In this
sense the proof presented in formulas
(\ref{rescmm})-(\ref{statabs}) does not differ from that of
\cite{Malec,Malec1}. Moreover, as it was noted in \cite{Manton},
in principle one can get an infinite number of integral identities
using the equations of motion (or local identities such as the
conservation of the energy-momentum tensor which follow from the
equations of motion). But it seems that considering the direct
transformations of the action under rescalings of the fields is
technically simpler at least for obtaining no-go results.

Now let us consider several examples. First, let us take the
simplest polynomial potential of the form
$V(\phi^{\dagger}\phi)=q(\phi^{\dagger}\phi)^{n}$ with $q>0$ and
$n\ge 1$. From (\ref{color1}) (which is nothing but (\ref{theor1})
with $\gamma=\frac{3}{2}$) we get
$(n-1)(\phi^{\dagger}\phi)^{n}\ge 0$ and thus such a potential
does not provide the existence of a non-topological soliton
solution satisfying (\ref{sol1}) and (\ref{sol2}). Second, let us
take a more complicated potential of the form
$V(\phi^{\dagger}\phi)=q_{1}(\phi^{\dagger}\phi)^{2}-q_{2}\phi^{\dagger}\phi$
with $q_{1}>0$, $q_{2}>0$ (this is the Higgs potential up to a
constant). Substituting this potential into (\ref{color1}) (of
course, we can use (\ref{color1}) not only for non-negative
potentials) we get $q_{1}(\phi^{\dagger}\phi)^{2}\ge 0$ and thus
this potential does not provide the existence of a non-topological
soliton solution satisfying (\ref{sol1}) and (\ref{sol2}) (we can
also take (\ref{theor1}) with, for example, $\gamma=1$ and get
$q_{1}(\phi^{\dagger}\phi)^{2}+q_{2}\phi^{\dagger}\phi\ge 0$,
which leads to the same conclusion). Meanwhile, for the "opposite"
potential
$V(\phi^{\dagger}\phi)=q_{1}\phi^{\dagger}\phi-q_{2}(\phi^{\dagger}\phi)^{2}$
with $q_{1}>0$, $q_{2}>0$ there is no $\gamma:\,
\frac{1}{2}<\gamma\le\frac{3}{2}$, for which (\ref{theor1}) holds
for any $\phi$ and thus there is no restriction for the existence
of solitons. Moreover, at least in the absence of the gauge field
a soliton is shown to exist in this case \cite{Derrick2}. The same
is valid for the potential
$V(\phi^{\dagger}\phi)=-q_{1}\phi^{\dagger}\phi\ln(q_{2}\phi^{\dagger}\phi)$
with $q_{1}>0$, $q_{2}>0$ -- there is no $\gamma:\,
\frac{1}{2}<\gamma\le\frac{3}{2}$, for which (\ref{theor1}) holds
for any $\phi$, and again at least in the absence of the gauge
field a soliton is shown to exist in this case \cite{Rosen4}. And
finally, let us take a potential of the form
$V(\phi^{\dagger}\phi)=q_{1}(\phi^{\dagger}\phi)^{4}-q_{2}(\phi^{\dagger}\phi)^{2}$
with $q_{1}>0$, $q_{2}>0$. Substituting this potential into
(\ref{theor1}) and taking $\gamma=\frac{3}{4}$ we get
$q_{1}(\phi^{\dagger}\phi)^{4}\ge 0$ and thus this potential does
not provide the existence of a non-topological soliton solution
satisfying (\ref{sol1}) and (\ref{sol2}). It should be noted that
the cases
$V(\phi^{\dagger}\phi)=q_{1}(\phi^{\dagger}\phi)^{2}-q_{2}\phi^{\dagger}\phi$
and
$V(\phi^{\dagger}\phi)=q_{1}(\phi^{\dagger}\phi)^{4}-q_{2}(\phi^{\dagger}\phi)^{2}$
with $q_{1}>0$, $q_{2}>0$ can be considered only as illustrative
examples. Indeed, although $\phi\equiv 0$, $A_{\mu}\equiv 0$ is a
solution to the corresponding equations of motion and in principle
it can be considered as the "vacuum" from the mathematical point
of view, $\phi=0$ is the local maximum of the scalar field
potential and thus such a configuration is unstable and can not be
considered as the vacuum from the physical point of view. Even if
there were non-topological solitons in these cases, obviously they
would be unstable.

At the end of this section let us discuss why we restrict
ourselves to considering solutions periodic in time up to
translations and spatial rotations. Of course, it is reasonable to
examine such a type of solutions when considering solitons. But
there is another reason, which concerns the technical side of the
method used above. When we obtain equations of motions we suppose
that the variations of fields vanish at spatial and time infinity.
But the variations of the fields coming from
(\ref{resc1})-(\ref{resc3}) clearly do not tend to zero at time
infinity in the general case and we have a contradiction between
the form of the variations used to obtain equations of motion and
the form of the variations used to show the absence of some
solutions to these equations of motion (it should be noted that
there is another problem, which can occur when considering the
full action with $t\in(-\infty,\infty)$ -- the method described
above can not be used for infinite integrals; for periodic
solutions we can take the effective action of form
(\ref{effact001}), which is finite by construction). In principle,
in the case of non-vanishing variations there can arise additional
terms in the action coming from the surface terms at time infinity
(an analogous problem, but in the case of the Derrick theorem
\cite{Derrick} applied to a finite space-time domain, was
discussed in \cite{Adib}). Considering solutions periodic in time
up to a spatial rotation and a coordinate shift makes the surface
terms at $t=0$ and $t=T$, which arise when obtaining equations of
motion from effective action (\ref{effact1}) with the help of
variations of fields following from (\ref{resc1})-(\ref{resc3}),
be modulo equal and cancel each other (note, that if, from the
very beginning, one wants to consider solitons which are not at
rest, it is necessary to use more complicated transformations,
which can be obtained from (\ref{resc1})-(\ref{resc3}) with the
help of the corresponding Lorentz transformations, and, of course,
the effective action with $\tilde T$ instead of $T$, to get the
canceling surface terms and the same no-go results\footnote{For
example, in the simple case $\vec l=(l,0,0)$ for the scalar field
one should take transformation of the form
$\phi(t,x,y,z)\to\lambda^{\gamma}\phi\left(\frac{(1-\lambda
v^{2})t+(\lambda-1)vx}{1-v^{2}}\,, \frac{(\lambda
-v^{2})x+(1-\lambda)vt}{1-v^{2}}\,,\lambda y\,,\lambda z\right)$
with $v=l/{\tilde T}<1$ instead of (\ref{resc1}). The variation of
the field, following from such a transformation, possesses the
same periodicity properties as the field itself.}). The latter
ensures that the equations of motion obtained from the initial
action and those obtained from the effective action coincide, as
well as possible solutions. In this case the procedure of
obtaining the no-go results described in this section appears to
be consistent with the equations of motion coming from the
original action.

\section{Charged massive vector field}
The scaling arguments, used above for the theory describing
Yang-Mills field coupled to a non-linear scalar field, can be used
to show the absence of solitons in other gauge field theories. As
an illustrative example, let us consider the massive complex
vector field coupled to the electromagnetic field. The action of
this theory has the form

\begin{equation}\label{actvec}
S=\int
d^{4}x\left[-\frac{1}{2}\eta^{\mu\rho}\eta^{\nu\sigma}{W^{-}_{\mu\nu}}W_{\rho\sigma}^{+}+m^2\eta^{\mu\nu}W^{-}_{\mu}W_{\nu}^{+}-\frac{1}{4}F^{\mu\nu}F_{\mu\nu}\right]
\end{equation}
with $m\ne 0$, where
\begin{equation}
F_{\mu\nu}=\partial_{\mu}A_{\nu}-\partial_{\nu}A_{\mu},
\end{equation}
\begin{equation}
D_{\mu}W^{\pm}_{\nu}=\partial_{\mu}W^{\pm}_{\nu}\mp ie
A_{\mu}W^{\pm}_{\nu},
\end{equation}
\begin{equation}
W^{\pm}_{\mu\nu}=D_{\mu}W^{\pm}_{\nu}-D_{\nu}W^{\pm}_{\mu}.
\end{equation}
Again we suppose that:
\begin{enumerate}
\item
there are no sources which are external to the system described by
action (\ref{actvec});
\item
all fields are smooth and vanish at spatial infinity;
\item
solutions to equations of motion are periodic in time up to a
spatial rotation and a coordinate shift (see previous section for
details).
\end{enumerate}

Let us denote (again in the coordinate system where $\vec l=0$)
\begin{eqnarray}\label{W1}
\int_{0}^{T}dt\int d^{3}x{W^{-}_{0i}}W_{0i}^{+}=\Pi_{W0}\ge 0,\\
\int_{0}^{T}dt\int
d^{3}x\frac{1}{2}{W^{-}_{ij}}W_{ij}^{+}=\Pi_{W1}\ge 0,
\end{eqnarray}
\begin{eqnarray}
m^2\int_{0}^{T}dt\int d^{3}x W^{-}_{0}W_{0}^{+}=V_{0}\ge 0,\\
m^2\int_{0}^{T}dt\int d^{3}x W^{-}_{i}W_{i}^{+}=V_{1}\ge 0,\\
\int_{0}^{T}dt\int d^{3}x \frac{1}{2}F_{0i}F_{0i}=\Pi_{A0}\ge 0,\\
\label{W2} \int_{0}^{T}dt\int d^{3}x
\frac{1}{4}F_{ij}F_{ij}=\Pi_{A1}\ge 0.
\end{eqnarray}
We also suppose that all these integrals are finite.

Analogously to what was made above, we also suppose that there
exists a solution to the corresponding equations of motion and
consider the following transformation of this solution:
\begin{eqnarray}
\label{resc0w}
W^{\pm}_{0}(t,\vec x)\to\lambda^{\beta-1}W^{\pm}_{0}(t,\lambda\vec x),\\
\label{resc1w}
W^{\pm}_{i}(t,\vec x)\to\lambda^{\beta}W^{\pm}_{i}(t,\lambda\vec x),\\
\label{resc2w} A_{0}^{a}(t,\vec x)\to A_{0}^{a}(t,\lambda\vec x),\\
\label{resc3w} A_{i}^{a}(t,\vec x)\to \lambda
A_{i}^{a}(t,\lambda\vec x)
\end{eqnarray}
with a real parameter $\lambda$. For the action on the transformed
fields we get
\begin{eqnarray}
S=\lambda^{2\beta-3}\Pi_{W0}-\lambda^{2\beta-1}\Pi_{W1}+\lambda^{2\beta-5}V_{0}-\\
\nonumber-\lambda^{2\beta-3}V_{1}+\lambda^{-1}\Pi_{A0}-
\lambda\Pi_{A1}.
\end{eqnarray}
Now let us take
\begin{equation}
\beta=\frac{3}{2}.
\end{equation}
We get
\begin{eqnarray}\label{variation1}
\frac{dS}{d\lambda}|_{\lambda=1}=-2\Pi_{W1}-2V_{0}-\Pi_{A0}-
\Pi_{A1}=0.
\end{eqnarray}
From this equation it follows that
\begin{eqnarray}
\Pi_{W1}=V_{0}=\Pi_{A0}=\Pi_{A1}\equiv 0,
\end{eqnarray}
which means that $F_{\mu\nu}\equiv 0$ and we can set
$A_{\mu}\equiv 0$, we also get $W^{\pm}_{0}\equiv 0$ and
$W^{\pm}_{ij}\equiv 0$. With $A_{\mu}\equiv 0$ we can rewrite
$W^{\pm}_{ij}\equiv 0$ as
\begin{equation}\label{firstc}
\partial_{i}W^{\pm}_{j}-\partial_{j}W^{\pm}_{i}\equiv 0.
\end{equation}

Now let us take the equations of motion for the field
$W^{\pm}_{\mu}$ with $A_{\mu}\equiv 0$. It follows from these
equations that
\begin{equation}
\partial^{\mu}W^{\pm}_{\mu}=0.
\end{equation}
Using the fact that $W^{\pm}_{0}\equiv 0$ we get
\begin{equation}\label{secondc}
\partial^{i}W^{\pm}_{i}=0.
\end{equation}
Equations (\ref{firstc}) and (\ref{secondc}) can be rewritten as
\begin{equation}\label{divrot}
\textrm{div}\,{\vec W}^{\pm}=0,\qquad \textrm{rot}\,{\vec
W}^{\pm}=0,
\end{equation}
where ${\vec W}^{\pm}=(W^{\pm}_1,W^{\pm}_2,W^{\pm}_3)$. Equations
(\ref{divrot}) imply that
\begin{eqnarray}
{\vec W}^{\pm}=\textrm{grad}\,\varphi^{\pm},\\
\Delta\varphi^{\pm}=0,
\end{eqnarray}
where $\varphi^{\pm}=\varphi^{\pm}(t,\vec x)$,
$(\varphi^{+})^{*}=\varphi^{-}$. The condition $\int d^{3}x
W^{-}_{i}W_{i}^{+}=\int d^{3}x
\partial_{i}\varphi^{-}\partial_{i}\varphi^{+}<\infty$ clearly leads to
$\varphi^{\pm}=\varphi^{\pm}(t)$ (one can simply use the results
of \cite{Hobart,Derrick} to show it) and therefore
\begin{equation}
{\vec W}^{\pm}\equiv 0.
\end{equation}
Thus
\begin{equation}
W^{\pm}_{\mu}\equiv 0.
\end{equation}
Finally, we can make the following statement:
\begin{prop}
Non-topological solitons, periodic in time up to a spatial
rotation and a coordinate shift such that $\Omega\vec l=\vec l$,
$|\vec l|/{\tilde T}<1$ and with finite integrals
(\ref{W1})-(\ref{W2}), are absent in the theory described by
action (\ref{actvec}).
\end{prop}

\section*{Acknowledgements}
The author is grateful to I.P. Volobuev for valuable discussions
and to an unknown referee of Journal of Mathematical Physics for
suggesting to consider rotations in addition to translations. The
work was supported by FASI state contract 02.740.11.0244 and by
grant of Russian Ministry of Education and Science NS-4142.2010.2.

\end{document}